\documentclass[apjl]{emulateapj}

\shorttitle{On the contribution of the X-ray source to the nebular HeII emission in IZw18}
\shortauthors{Carolina Kehrig et al.}

\usepackage[usenames]{color}

\begin{document}

\title{On the contribution of the X-ray source to the extended nebular HeII emission in IZw18}

\author{C. Kehrig\altaffilmark{1}}
\author{M.A. Guerrero\altaffilmark{1}}
\author{J. M. V\'\i lchez\altaffilmark{1}}
\author{G. Ramos-Larios\altaffilmark{2}}

\affil{\altaffilmark{1}
Instituto de Astrof\'\i sica de Andaluc\'\i a (IAA-CSIC), Glorieta de la Astronom\'\i a s/n, E-18008 Granada, Spain}
\affil{\altaffilmark{2}
Instituto de Astronom\'\i a y Meteorolog\'\i a, Dpto. de F\'\i sica, CUCEI, Universidad de Guadalajara, Av. Vallarta No. 2602, C.P. 44130, Guadalajara, Jalisco, Mexico}

\begin{abstract}
Nebular He{\sc ii} emission implies the presence of energetic photons
(E$\ge$54 eV). Despite the great deal of effort dedicated to understanding He{\sc ii}
ionization, its origin has remained mysterious, particularly in
metal-deficient star-forming (SF) galaxies. Unfolding He{\sc ii}-emitting,
metal-poor starbursts at z$\sim$0 can yield insight into the powerful
ionization processes occurring in the primordial universe. Here we
present a new study on the effects that X-ray sources have on the He{\sc ii}
ionization in the extremely metal-poor galaxy IZw18 
(Z $\sim$3$\%$ Z$_{\odot}$), whose
X-ray emission is dominated by a single high-mass X-ray binary
(HMXB). This study uses optical integral field spectroscopy,
archival Hubble Space Telescope observations, and all of the X-ray
data sets publicly
available for IZw18. 
We investigate the time-variability of the IZw18 HMXB for the first time; 
its emission shows small variations on timescales from days to decades. 
The best-fit models for the HMXB X-ray spectra cannot reproduce the observed 
He{\sc ii} ionization budget of IZw18, nor can recent photoionization models that combine the spectra of both very low metallicity massive stars and the emission from HMXB.
We also find that the IZw18 HMXB and the He{\sc ii}-emission peak are 
spatially displaced at a projected distance of $\simeq$ 200 pc. 
These results reduce the relevance of X-ray photons as the dominant He{\sc ii} 
ionizing mode in IZw18, which leaves uncertain what process is
responsible for the bulk of its HeII ionization.
This is in line with recent work discarding X-ray binaries as the main source 
responsible for He{\sc ii} ionization in SF galaxies.
\end{abstract}


\section{Introduction}

\label{sec:Intro}

The presence of nebular He{\sc ii} emission points to the existence
of a hard radiation field as photon energies $\ge$ 54 eV are necessary
to ionize He$^{+}$. Highly ionized systems, like He{\sc ii} emitters,
are expected to be more frequent at high-z
\citep[e.g.,][]{cassata,mainali}. At the same time, observations
of local, He{\sc ii}-emitting star-forming (SF) galaxies suggest that the lower the metallicity (Z), the larger the nebular
He{\sc ii} line intensities \citep[e.g.,][]{G00,sen17}. This is in
line with the harder ionizing stellar spectra predicted at the lower
metallicities typical in the early universe \citep[e.g.,][]{TS00};
the first, hot metal-free Population III stars produce
strong nebular He{\sc ii} lines
\citep[e.g.,][]{Y12,visbal15}. Still, despite much
observational and theoretical effort, including relevant advances in
stellar modeling, the source of He{\sc ii} ionization remains puzzling, especially in metal-poor galaxies
\citep[e.g.,][]{bpass,go18,K18,se19,kub19}.

There is evidence for enhanced total X-ray luminosities per unit of
star formation rate ($L_{\rm X}$/SFR) and high-mass X-ray binary
(HMXB) populations in extremely metal-poor galaxies (Z/Z$_{\odot}$ $<$
0.1) compared to those of $\sim$Z$_{\odot}$
\citep[e.g.,][]{pre13,bro16,l17,S19,ponada}. Furthermore, examples of
nearby He{\sc ii} nebulae associated with HMXBs that appear to act as
the only ionizing source are reported
\citep[e.g.,][]{G91,kaaret04,gu14}. Thus HMXBs have been proposed as
one of the leading mechanisms to ionize He$^{+}$ in SF systems.
However, it has been recently questioned whether they can provide the
required He$^+$ ionizing flux \citep[e.g.,][]{K18,sen20,saxena}.
Variability of the X-ray source could alleviate this conflict, but so
far only the extensive study of the puzzling He{\sc
  ii}$\lambda$4686-emitting nebula N44C in the LMC considered the
effects of an on-off switching X-ray source in time periods of years
\citep[see][]{naze}.

IZw18 is among the most metal-poor (12+log(O/H)= 7.11 $\sim$ 3$\%$
solar\footnote{ Assuming a solar metallicity 12+log(O/H)$_{\odot}$ =
  8.69 (Asplund et al.\ 2009)}; e.g., Kehrig et al.\ 2016, K16) He{\sc
  ii}-emitting SF galaxies known at z $\sim$ 0.  It thus can
be considered a nearby analog of distant metal-poor He{\sc ii}
emitters, providing a unique laboratory to study the high-ionization
phenomenon expected to be common in the reionization era.  Based on
integral field spectroscopy (IFS) data, \citet[][K15]{K15}
unveiled the entire nebular He{\sc ii}$\lambda$4686-emitting region in
IZw18 with a diameter of $\approx$5$^{\prime\prime}$ ($\approx$ 440 pc
at the distance of 18.2 Mpc; Aloisi et al. 2007). K15 found that only
peculiar hot (nearly) metal-free massive stars could explain the
He{\sc ii} ionization in IZw18 \citep[see also][K18]{K18}.  
The main goal of this Letter is to investigate in detail the spatial correlation between the He{\sc ii} 
emission in IZw18 and its single HMXB \citep[][T04]{Thuan2004}, 
the variability of the latter on short (days) and long (decades) 
timescales and its expected He$^+$ ionizing photon production to assess the contribution of the X-ray emission to the 
nebular He{\sc ii}$\lambda$4686 ionization.


\begin{table*}\centering
\setlength{\columnwidth}{0.1\columnwidth}
\setlength{\tabcolsep}{1.0\tabcolsep}
\caption{X-ray Observations of IZw18 \label{tab.xray.obs}}
\begin{tabular}{lllcrc}
\hline
\hline

\multicolumn{1}{l}{X-ray Observatory} &
\multicolumn{1}{c}{Instrument} & 
\multicolumn{1}{c}{Data set} &
\multicolumn{1}{c}{Date} & 
\multicolumn{1}{c}{Net Exposure Time} & 
\multicolumn{1}{c}{Count Rate$^a$} \\

\multicolumn{1}{l}{} & 
\multicolumn{1}{c}{} & 
\multicolumn{1}{c}{} & 
\multicolumn{1}{c}{} & 
\multicolumn{1}{c}{(ks)} & 
\multicolumn{1}{c}{(ks$^{-1}$)} \\
\hline

ROSAT      & PSPC       & rp600165n00 & 1992 04 30 &       16.2~~~~~~~~~~ &   7.2$\pm$0.7 \\
Chandra    & ACIS-I     & 600108      & 2000 02 08 &       31.0~~~~~~~~~~ &  12.3$\pm$0.6 \\
XMM-Newton & EPIC-pn    & 0112520101  & 2002 04 10 &       24.2~~~~~~~~~~ &  90.4$\pm$2.0 \\
                  & EPIC-MOS1  &             &            &       31.1~~~~~~~~~~ &  28.3$\pm$1.0 \\
                  & EPIC-MOS2  &             &            &       30.9~~~~~~~~~~ &  28.4$\pm$1.0 \\
XMM-Newton & EPIC-pn    & 0112520201  & 2002 04 16 &        5.4~~~~~~~~~~ &  76.9$\pm$4.0 \\
                  & EPIC-MOS1  &             &            &       10.4~~~~~~~~~~ &  26.8$\pm$1.7 \\
                  & EPIC-MOS2  &             &            &       10.4~~~~~~~~~~ &  29.2$\pm$1.7 \\
Swift     & XRT        & 00040765001 & 2011 03 25 &        2.0~~~~~~~~~~ &   2.1$\pm$1.1 \\
Swift     & XRT        & 00040765002 & 2011 03 29 &        0.9~~~~~~~~~~ &   6.6$\pm$2.8 \\
Swift     & XRT        & 00040765003 & 2012 03 23 &        0.4~~~~~~~~~~ &   9.2$\pm$5.0 \\
Swift     & XRT        & 00040765004 & 2012 03 26 &        0.3~~~~~~~~~~ &   8.9$\pm$5.6 \\
Swift     & XRT        & 00040765005 & 2012 08 29 &        2.6~~~~~~~~~~ &   6.7$\pm$1.6 \\

\hline
\end{tabular}
\begin{flushleft}
{(a) Count rates are computed within the nominal energy bands of each X-ray observatory: ROSAT/PSPC (0.1-2.4 keV), Chandra/ACIS-I (0.3-10.0 keV), XMM-Newton/EPIC (0.3-10.0 keV), and Swift/XRT (0.2-10.0 keV).} 
\end{flushleft}
\end{table*}

\section{Observations}

This study combines archival X-ray and Hubble Space Telescope (HST) observations with IFS data.

The IFS observations of IZw18, obtained on 2012 December, use the Potsdam Multi-Aperture
Spectrophotometer \citep[PMAS;][]{roth05} at the Calar Alto Observatory
 (Spain). PMAS provides a field
of view (FOV) of $16^{\prime\prime}\times16^{\prime\prime}$ with a sampling
of  $1^{\prime\prime}\times1^{\prime\prime}$ over
$\sim$3640--7200 \AA~(see K15,K16 for details).  
The data have been used to obtain a continuum-subtracted spectral map
of the nebular He{\sc ii}$\lambda$4686 emission line in IZw18.  

IZw18 has been targeted by most modern X-ray observatories, including
ROSAT, Chandra, XMM-Newton and Swift.  Table~\ref{tab.xray.obs} lists
the details of these X-ray observations.  The Chandra and XMM-Newton
data were reprocessed using the specific reduction packages Chandra
Interactive Analysis of Observations \citep[CIAO
v4.9,][]{Fruscione2006} and XMM-Newton Science Analysis System
\citep[SAS v17.0,][]{Gabriel2004}, respectively.  The ROSAT and
Swiftdatasets were processed using generic HEASARC reduction
packages. IZw18 is detected as a point source in all these observations
except in the Chandra observations, with the highest spatial resolution
($\approx$0\farcs5) among the available X-ray data, which suggest
faint extended emission toward the south
(T04). Table~\ref{tab.xray.obs} provides the count rates of these
detections, which are below a 3$\sigma$ threshold in the shortest
Swiftobservations.

We also use archival HST images of IZw18.  The location of the hot,
young massive stars is probed using the HST STIS image obtained
through the far-UV filter F25SRF2 with $\lambda_c$=1457 \AA\
\citep[PI:T.M.\ Brown, prop.\ 9054,][]{BR02}. The HST WFPC2/F658N
image \citep[PI:E.\ Skillman, prop.\ 6536,][]{D98} outlines the
ionized gas emission in the H$\alpha$ line.

\section{Data Analysis}

\subsection{UV, optical and X-ray Imaging}

Previous analyses of the spatial properties of the X-ray emission in
IZw18 were presented by T04 using the same Chandra data set used
here. They suggested the presence of an HMXB spatially coincident with
the bright northwest (NW) SF knot of IZw18 and a fainter diffuse
emission toward the south.  To investigate the spatial distribution of
the X-ray emission from IZw18, we have produced images in the soft
0.3-1.2 keV, medium 1.2-2.5 keV, hard 2.5-8.0 keV, and broad 0.3-8.0
keV X-ray bands using the ACIS subpixel event resolution capability,
with subpixel size 1/4 their natural size of 0\farcs5.  Images were
subsequently convolved with subpixel PSF images of the same energy
and adaptively smoothed with a Gaussian kernel size between
0\farcs5-2\farcs0.

To spatially correlate the different components of IZw18, we compare
in detail the distribution of the emission of X-ray with that of the
hot young massive stars and ionized gas from HST images, and with the
PMAS He{\sc ii}$\lambda$4686 map. The HST and Chandra images were
registered using USNO stars in their FoV. As noted by T04, the
astrometric accuracy of the Chandra observations is $\approx$0\farcs5
due to the limited number of optical counterparts of X-ray sources.
Similarly, the HST images and PMAS He{\sc ii}$\lambda$4686 map were
registered using the HST image and PMAS map in the H$\alpha$ emission line. Their relative registration accuracy is estimated to be within
$\approx$0\farcs5.

\begin{figure*}
\centering
\includegraphics[bb=1 1 768 768,width=0.33\textwidth]{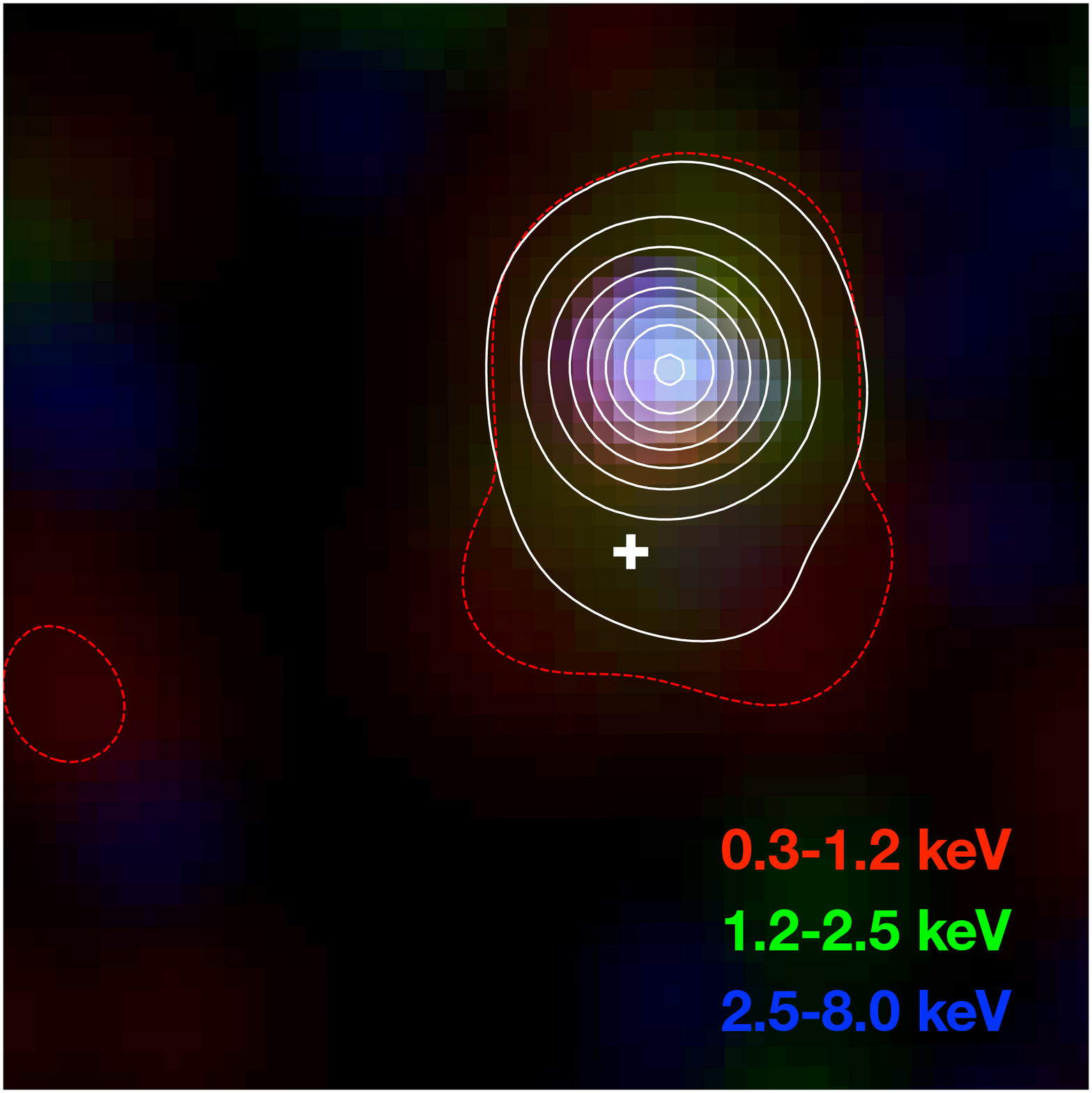}
\includegraphics[bb=1 1 768 768,width=0.33\textwidth]{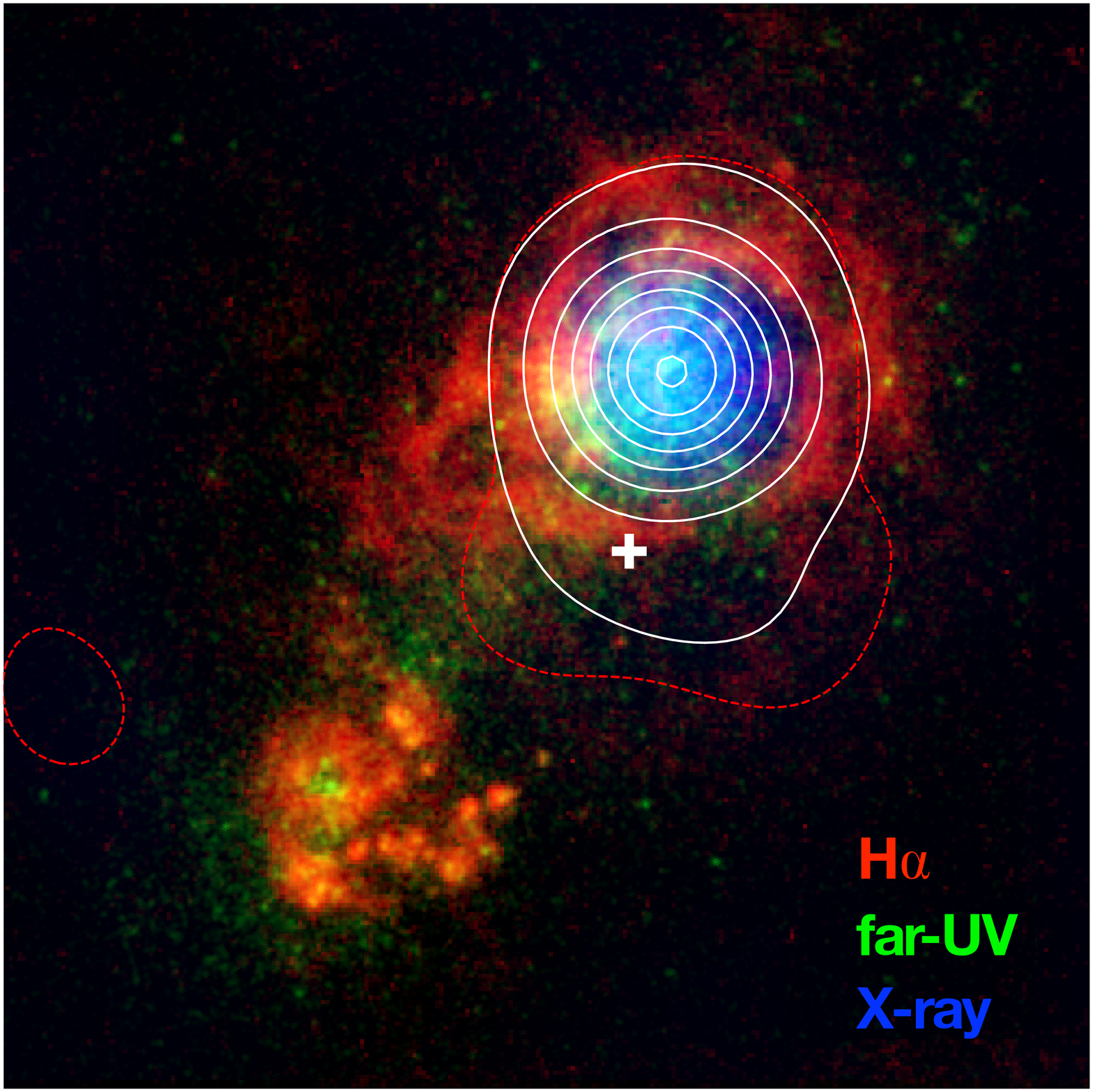}
\includegraphics[bb=1 1 768 768,width=0.33\textwidth]{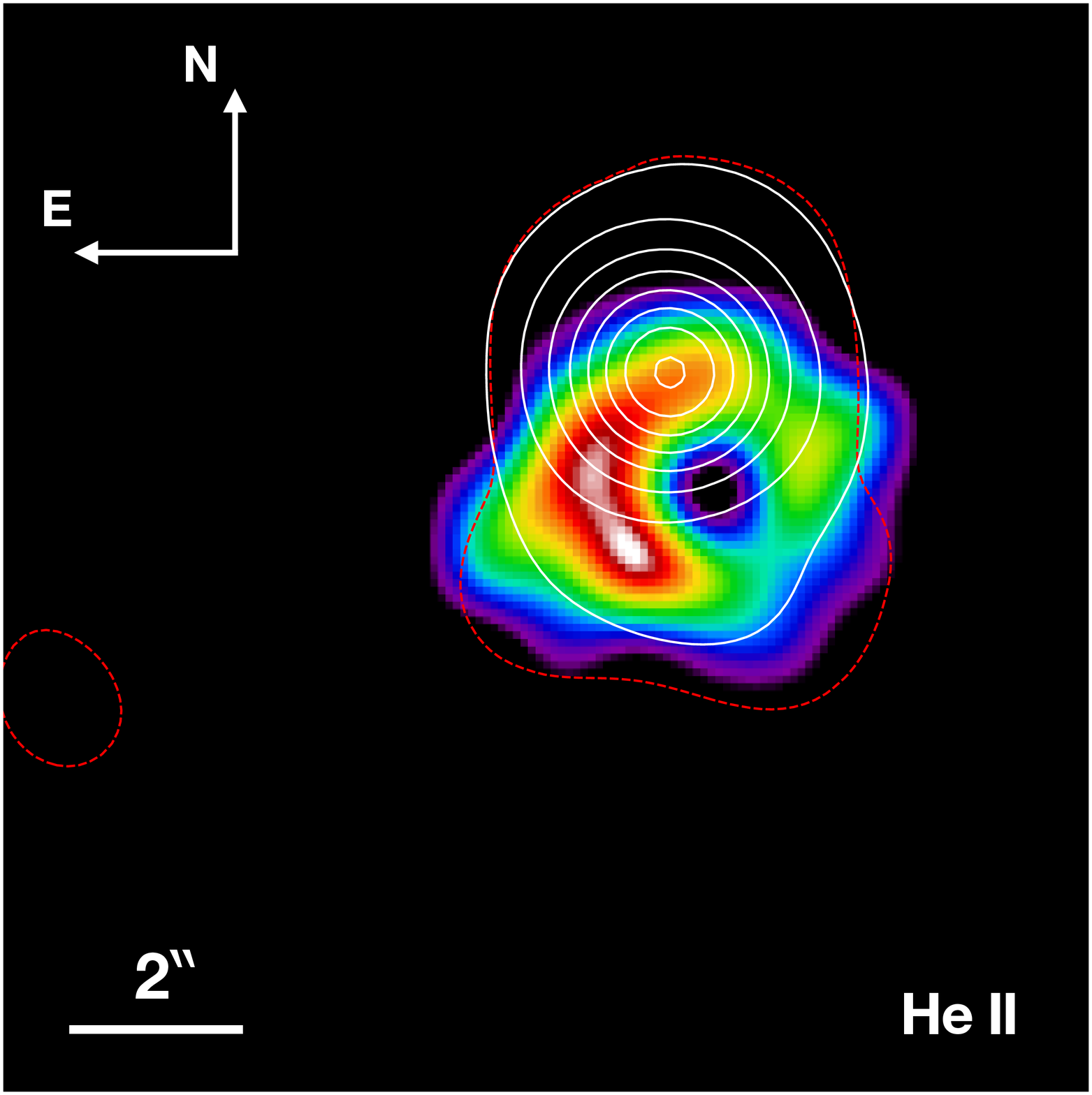}
\caption{Multiwavelength images of IZw18 and Chandra ACIS-I X-ray contours.
(Left panel) 
Color-composite Chandra ACIS-I X-ray picture in the soft 0.3-1.2 keV (red), medium 1.2-2.5 keV (green), and hard 2.5-8.0 keV (blue) energy bands.
(Center panel) 
Color-composite optical HST WFPC2/F658N H$\alpha$ emission line (red), HST STIS/FUV/F25SRF2 continuum at 1457 \AA\ (green), and X-ray Chandra ACIS-I 0.3-8.0 keV (blue) picture. 
(Right panel)
PMAS map of nebular He{\sc ii}$\lambda$4686 emission. 
In this map, the pixel scale has been regrid to 0\farcs1 applying a 2D
linear interpolation, and then
smoothed at the original pixel scale using a
$\sigma$=1$^{\prime\prime}$ Gaussian kernel; intensity is presented in
the square root scale.
The orientation, FoV, and scale of the three panels match each other.
The He{\sc ii} emission peak in the right panel is marked by a '+'
symbol in the left and central panels. The white contours correspond to the PSF of the X-ray point
source, and display the intensity of the 0.3-8.0 keV X-ray emission
using a logarithmic scale from a 3$\sigma$ level to the intensity
peak; the red contour shows the 3$\sigma$ level of the soft 0.3-1.2 keV emission.
The nominal FWHM of the Chandra ACIS PSF, $\simeq$0\farcs5, is
in between the first and second highest white contours, whereas the He{\sc ii}
emission peak is observed well outside the latter.}
\label{fig.img}
\end{figure*}

Figure~\ref{fig.img} (left) presents a color-composite picture of the soft, medium, and
hard X-ray band images. This image confirms the presence of a hard X-ray point source at the
position of the NW knot (T04) and reveals 
that the diffuse emission toward the south is redder, i.e.,
the diffuse emission is softer than the point source.
Figure~\ref{fig.img} (center) shows that the HMXB emission (blue)
is coincident with the young massive stellar cluster
(green) centered in the NW component of H$\alpha$ emission (red) of IZw18.
The diffuse X-ray emission extends south, along some holes in the H$\alpha$ emission.
Figure~\ref{fig.img} (right) displays the extended nebular He{\sc
  ii}$\lambda4686$-emitting gas showing, for the first time, that the He{\sc ii}$\lambda4686$ emission
peak is displaced $\approx$2\farcs5 ($\approx$200 pc projected at the distance of 18.2
Mpc) from the south of the HMXB in IZw18. The He{\sc ii} emission
overlaps part of the much more extended H$\alpha$ emission, whose
peaks appear noncoincident (see also Figure 2 (middle) panels from K15).

\subsection{X-ray Variability}
\begin{figure*}
\centering
\includegraphics[bb=18 160 590 718,width=0.95\textwidth]{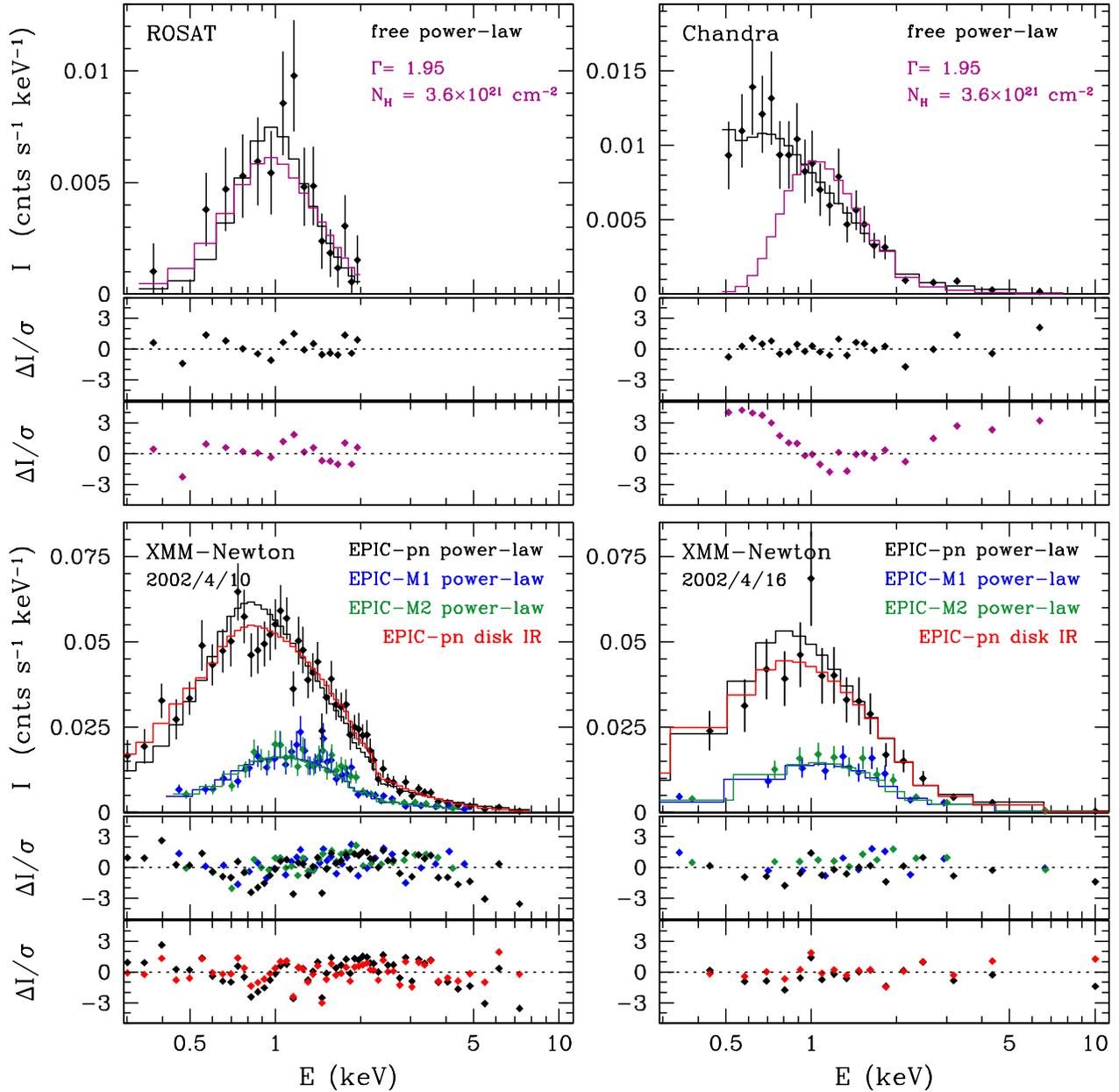}
\caption{
Background-subtracted ROSAT (top left), Chandra 
(top right), and XMM-Newton (bottom) X-ray
spectra of IZw18 (points) overplotted by best-fit models (histograms).
The error bars of each data point correspond to 1$\sigma$ statistical
uncertainties.  
The best-quality XMM-Newton spectra are fitted using an absorbed power
law
(black for EPIC-pn, blue for EPIC-MOS1, and green for EPIC-MOS2).
An absorbed irradiated disk
model is also shown for the EPIC-pn spectrum
(red).  
The ROSAT and Chandra spectra are fitted using an absorbed
power law models (black), with the scaled best-fit power law model of
the 2002-04-10-XMM-Newton observations also overplotted (purple).
Lower panels show the residuals of the best-fit models where the data points' colors correspond to those of the
best-fit models.
\label{fig.xray.spec}}
\end{figure*}

We used the different X-ray datasets listed in Table~\ref{tab.xray.obs} to
extract X-ray spectra utilizing apertures encompassing the X-ray emission
associated with the NW SF knot. 
Background spectra were extracted from nearby source-free regions. The 
source and background spectra were then processed to obtain calibration
files.
The background-subtracted ROSAT/PSPC,
Chandra/ACIS-I, and XMM-Newton/EPIC X-ray spectra of
IZw18 are shown in Figure~\ref{fig.xray.spec}.
The Swiftspectra are not shown given their limited count number.

Previous analyses of the X-ray spectrum of the IZw18 HMXB considered 
plasma emission and power-law models for ROSAT \citep{Martin1996}
and Chandra data (T04), and power-law and thin accretion
disk around a Kerr black hole (BH) models for the XMM-Newton data
obtained on 2002 April 10 \citep[2002-04-10-XMM-Newton;][]{KF2013}. 
The latter model is the most specific for HMXBs, but too complex
for the quality of some of the available datasets.  
To preserve the consistency of the investigation of the HMXB X-ray variability, a power-law model will be used for all datasets.
For the fit, a fixed Galactic extinction along the line of sight of IZw18
consistent with a hydrogen column density ($N_{\rm H}$) of
2.9$\times$10$^{20}$ cm$^{-2}$ \citep{HI4PI} is assumed.
An additional absorption component with the same chemical abundance as IZw18 is
adopted to describe the local absorption.

The results of our spectral modeling are summarized in Table~\ref{tab.xray.model}.
We first started modeling the 2002-04-10-XMM-Newton spectra, which have
the largest count number.
The power-law model (black, blue, and green histograms in
Figure~\ref{fig.xray.spec}) results in an acceptable fitting ($\chi^2$/DoF=1.33).
The best-fit power-law photon index $\Gamma$=1.95 is typical of
synchrotron emission, whereas the $N_{\rm H}$ of the local component
(=3.6$\times$10$^{21}$ cm$^{-2}$) is consistent with that derived by
\citet{Lelli2012} at the location of the X-ray source.
We then fitted the XMM-Newton/EPIC spectra extracted from the
observations on 2002 April 16, using a similar absorbed power-law model.
The spectra have lower quality than those obtained just 6 days before given
their shorter net exposure times, but the parameters of the best-fit models
are consistent within their uncertainties with those obtained for the 2002-04-10-XMM-Newton
observations (see Table~\ref{tab.xray.model}).

\begin{table*}\centering
\setlength{\columnwidth}{0.1\columnwidth}
\setlength{\tabcolsep}{1.0\tabcolsep}
\caption{X-ray Spectral Fittings and Luminosities of IZw18 \label{tab.xray.model}}
\begin{tabular}{lcccrcccc}
\hline
\hline

\multicolumn{1}{l}{X-ray Observatory} &
\multicolumn{1}{c}{Date} & 
\multicolumn{1}{c}{N$_{\rm H}$} & 
\multicolumn{1}{c}{$\Gamma$} &
\multicolumn{1}{c}{$\chi^2$/DoF} & 
\multicolumn{1}{c}{$f_{\rm 0.3-2.4 keV}$} & 
\multicolumn{1}{c}{$f_{\rm 0.5-8.0 keV}$} & 
\multicolumn{1}{c}{$L_{\rm 0.3-2.4 keV}$} &
\multicolumn{1}{c}{$L_{\rm 0.5-8.0 keV}$} \\

\multicolumn{1}{l}{and Instrument} & 
\multicolumn{1}{l}{} & 
\multicolumn{1}{l}{(10$^{21}$ cm$^{-2}$)} & 
\multicolumn{1}{l}{} & 
\multicolumn{1}{l}{} & 
\multicolumn{2}{c}{(10$^{-14}$ erg~cm$^{-2}$~s$^{-1}$)} & 
\multicolumn{2}{c}{(10$^{38}$ erg~cm~s$^{-1}$)} \\
\hline

ROSAT PSPC$^{a}$       & 1992 04 30 & 16$^{+34}_{-5}$       & 4.2$^{+4.3}_{-2.0}$ & 12.7/14 = 0.91 &            3.8         &      6.5           & 400$^{+20}_{-150}$    & 127$^{+6}_{-60}$  \\
                        &            & 3.6                  & 1.95              & 16.5/16 = 1.03 &            7.5         &      8.9           & 30.0$^{+4.2}_{-3.3}$ &  35.5$^{3.8}_{-4.4}$ \\
Chandra ACIS-I   & 2000 02 08 & 0.5$^{+1.3}_{-0.5}$   & 1.96$^{+0.28}_{-0.21}$ & 15.2/20 = 0.76   &        3.7         &      6.3            & 19.4$^{+0.8}_{-1.5}$ &  26.9$\pm$2.1       \\
XMM-Newton EPIC  & 2002 04 10 & 3.6$\pm$0.6         & 1.95$\pm$0.07        & 158.7/122 = 1.33 &        13.5        &     28.0            &  94.3$\pm$1.5      & 131.2$^{+1.9}_{-4.0}$ \\
XMM-Newton EPIC  & 2002 04 16 & 3.2$^{+1.1}_{-0.9}$   & 1.84$\pm$0.13        & 33.6/38 = 0.88   &        12.1        &     26.8            &  48.4$\pm$2.1      & 107.2$\pm$4.4       \\
Swift/XRT        & 2011 03 25 & 3.6                  & 1.95              & $\dots$            &         4.5        &       9.2           &  31$\pm$17         &  44$\pm$23          \\
Swift/XRT        & 2011 03 29 & 3.6                  & 1.95              & $\dots$            &        14          &      29             &  98$\pm$42         & 135$\pm$60          \\
Swift/XRT        & 2012 03 23 & 3.6                  & 1.95              & $\dots$            &        20          &      41             & 135$\pm$70         & 190$\pm$100         \\
Swift/XRT        & 2012 03 26 & 3.6                  & 1.95              & $\dots$            &        19          &      39             & 130$\pm$80         & 180$\pm$110         \\
Swift/XRT        & 2012 08 29 & 3.6                  & 1.95              & $\dots$            &        14          &      30             &  98$\pm$23         & 138$\pm$33          \\

\hline
\end{tabular}
\begin{flushleft}
(a) First and second rows show the results of ROSAT spectral modeling based on a free power-law fit and using the best-fit model of the 2002-04-10-XMM-Newton spectrum, respectively. 
\end{flushleft}
\end{table*} 

\begin{figure*}
\centering
\includegraphics[bb=18 255 590 710,width=0.75\textwidth]{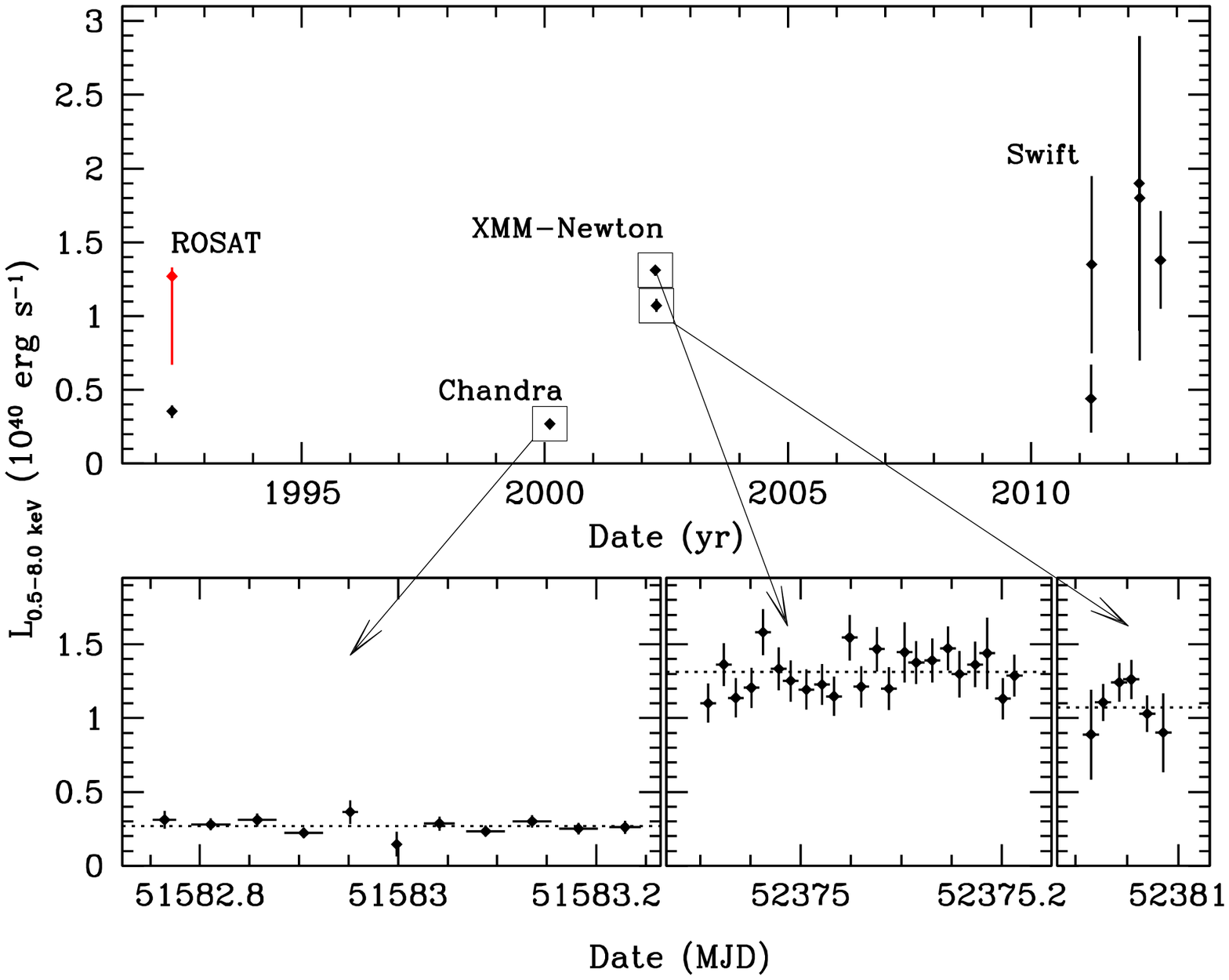}
\caption{Long-term (top) and short-term (bottom) X-ray variability of the X-ray emission from IZw18.  
The intrinsic $L_{\rm X,0.5-8~keV}$ and corresponding error bars are derived from the models detailed in Table~2. The ROSAT red data point corresponds to a free power-law fit, whereas the black 
data point uses the best-fit parameters of the fit to the
2002-04-10-XMM-Newton spectrum. The bottom panels present expanded light-curves of the Chandra and XMM-Newton 
observations.  
The dashed horizontal lines in these panels correspond to the average X-ray luminosities 
during the observations.
\label{fig.xray.lc}}
\end{figure*}

Next we proceed to fit the ROSAT/PSPC and Chandra/ACIS-I 
spectra with this same absorbed power-law model.
The parameters of the best-fit models (black histograms in
the top panels of Figure~\ref{fig.xray.spec}) are notably different from
those obtained for the XMM-Newton/EPIC spectra.  As for the
ROSAT/PSPC spectrum, the data quality is not adequate to reliably
constrain the model parameters (see their 1$\sigma$
uncertainties in Tab.~\ref{tab.xray.model}) and indeed the scaled
best-fit model of the 2002-04-10-XMM-Newton spectra (purple
histogram in the top left panel of Figure~\ref{fig.xray.spec}) is
consistent with the ROSAT/PSPC spectrum.  On the other hand,
the Chandra ACIS-I spectrum is much softer than the scaled
best-fit model of the 2002-04-10-XMM-Newton spectra (purple
histogram in the top right panel of Figure~\ref{fig.xray.spec}), i.e.,
the internal extinction during the Chandra observation was much
lower than that during the XMM-Newton observations.  The
Chandra spectrum also implies the lowest $L_{\rm X}$. Finally,
we have adopted the best-fit model of the 2002-04-10-XMM-Newton
spectra to derive the $L_{\rm X}$ from the count rate of the
Swift observations.  These are listed in
Table~\ref{tab.xray.model}, where the uncertainties are derived from
the 1$\sigma$ error bar of the count rate.

The long-term evolution of $L_{\rm X}$ in the energy
band 0.5-8.0 keV ($L_{\rm X,0.5-8~keV}$; see Table~\ref{tab.xray.model}) is shown in the upper panel of
Figure~\ref{fig.xray.lc}.  
Apparently, the IZw18 HMXB switches from a low state with $L_{\rm X,0.5-8~keV}$ 
a few times 10$^{39}$ erg~s$^{-1}$ to an upper state with $L_{\rm X,0.5-8~keV}$ 
of (1--2)$\times$10$^{40}$ erg~s$^{-1}$.  
These high $L_{\rm X}$ and spectral changes are consistent with those 
of ultraluminous X-ray sources, as discussed by \citet{KF2013}, although 
T04 argued that HMXBs in low metallicity galaxies tend to 
have higher $L_{\rm X}$.
The transition from one state
to the other occurs in timescales shorter than 2-3 yr, as suggested
by the increase in luminosity from the Chandra to the
XMM-Newton observations, and from the different Swift
observations, although the latter are affected by large uncertainties
in the luminosity estimates.

The short-term variability is shown in the lower panels of
Figure~\ref{fig.xray.lc}.
Lightcurves in the energy band 0.5-8.0 keV with a bin size of 4.0 and 1.2
ks have been extracted from the Chandra/ACIS-I and XMM-Newton/EPIC-pn observations, respectively.
The count rates and their 1$\sigma$ uncertainties have been converted
to X-ray luminosities using the average $L_{\rm X,0.5-8~keV}$ at each epoch given in the last column of
Table~\ref{tab.xray.model}.
For each data set, the X-ray emission is mostly consistent with its
mean value, which may be suggesting small fluctuations.
In this sense, the X-ray level of the two XMM-Newton observations,
obtained $\sim$6 days apart, are mostly consistent with each other.  

This leads us to conclude that the IZw18 HMXB has a
relatively constant emission in short (hours to days) timescales,
with variations of factors of only a few in longer (a few years) timescales.
There are hints that the larger X-ray fluxes are associated with
an increase of $N_{\rm H}$ toward this source, as
typically seen in X-ray sources fed by accretion such as HMXBs. 

\section{Discussion}

The nebular He{\sc ii}$\lambda$4686 emission observed in some SF
galaxies has often been attributed to X-ray emission from HMXBs
\citep[e.g.][]{TI05,S19}. This hypothesis has been critically
tested by \citet[][S20]{sen20} using photoionization grids
accounting for the combined contribution of HMXBs and young massive
stars for metallicities from Z=0.02 to Z=0.001 ($\simeq$5$\%$
solar), and provide predictions for the He{\sc ii}/H$\beta$ ratio
that correspond to a range of values of the $L_{\rm X,0.5-8~keV}$/SFR
ratio.  We estimate a SFR of 0.14 M$_\odot$~yr$^{-1}$ for IZw18\footnote{The IZw18 integrated $F(H\alpha)$=4.36$\times$10$^{-13}$erg~s$^{-1}$~cm$^{-2}$ (K16) was translated to $L(H\alpha)$ assuming a distance of 18.2 Mpc.}
based on its $L(H\alpha)$=1.73$\times$10$^{40}$erg~s$^{-1}$(K16) and
the widely used {\it
  SFR-L(H$\alpha$)} relation from \cite{ken98}.  Therefore, taking the
$L_{\rm X,0.5-8~keV}$ value in Table~\ref{tab.xray.model} derived from
the 2002-04-10-XMM-Newton data set, this implies an observed
$L_{\rm X,0.5-8~keV}$/SFR of 9$\times$10$^{40}$
erg~s$^{-1}$/(M$_\odot$~yr$^{-1}$).  According to S20's
predictions, this $L_{\rm X,0.5-8~keV}$/SFR value results in He{\sc
  ii}/H$\beta$ between 0.0004 to 0.0025, regardless of the metallicity
assumed in the models, a value well below the observed He{\sc ii}/H$\beta$ of
0.023$\pm$0.004 in IZw18 (K16).
To reinforce this result, we have computed the total He{\sc ii}-ionizing photon
flux $Q$(He{\sc ii}) from Kerr BH and irradiated disk models of the X-ray
emission detected in the 2002-04-10-XMM-Newton observation.
The thin accretion disk around a Kerr BH model including a
Comptonization component provides an extrapolation to UV wavelengths
$<$ 228 \AA\ similar to that of the multicolor disk models
used by S20.
The best fit for this model has very similar parameters to those reported
by \citet{KF2013}, with an acceptable fitting quality ($\chi^2$/DoF=0.90).
The intrinsic (unabsorbed) $\log Q$(He{\sc ii}) for this model is
$\approx$48.4 photon~s$^{-1}$, i.e.,$\approx$50 times lower than
that of $\log Q$(He{\sc ii})=50.1 photon~s$^{-1}$ reported by K15.  
Meanwhile, the emission model of a disk irradiated by the Compton tail
\citep{GDP2009} is of particular interest, as the physical processes of
disk irradiation provide a meaningful description of the expected UV
spectrum to constrain more reliably $Q$(He{\sc ii}) \citep{BD2012}.  
The best fit for this model has a slightly better fit quality
($\chi^2$/DoF=0.93) with a local $N_{\rm H}$=1.5$\pm$0.8$\times$10$^{21}$ cm$^{-2}$, 
inner disk temperature 0.55$\pm$0.25 keV, 
photon index 2.9$\pm$0.6, and irradiated flux fraction of 2.3$\%$ for the 
\citep[other parameters were fixed at typical values according to;][]{GDP2009}.  
The resulting intrinsic (unabsorbed) $\log Q$(He{\sc ii}) is
$\approx$49.1 photon~s$^{-1}$, which is still 10 times lower than
observed.  
It is worthwhile noting that the $L_{\rm X}$ derived from
these models is very similar to that derived from a power-law model. 

Furthermore, the HMXB is coincident with the NW stellar cluster of
IZw18, whereas the nebular He{\sc ii} emission (diameter $\approx$440 pc)
shows a shell-like structure extending south of the HMXB; a $\approx$2.$^{\prime\prime}$5 (200 pc) offset is observed between
the He{\sc ii} emission peak and the HMXB (see Section 3.1).
Interestingly, the diffuse X-ray emission to
the south of the HMXB delineates the southernmost edge of the He{\sc
  ii} emission.  A spectral analysis of this diffuse emission is not
feasible, but its photon median energy, $\simeq$0.7 keV, is consistent
with that found in superbubbles \citep[e.g., DEM\,L152 in the
  LMC;][]{Jaskot_etal2011}.  The ``true'' spatial distribution of the
diffuse X-ray emission might certainly fill the cavity of ionized gas
suggested by its He{\sc ii} and H$\alpha$ emission, but
the bright X-ray emission from the HMXB prevents the detection of
its northern rim. Assuming that the X-ray-emitting plasma responsible of the diffuse 
emission had similar temperature as that of the typical superbubble DEM\,L152 
\citep[0.37 keV;][]{Jaskot_etal2011}, the count rate from the diffuse 
emission of 0.39$\pm$0.11 counts ks$^{-1}$ ($\sim$30 times smaller than that of the HMXB) implies $L_{\rm X,0.5-8~keV}$ $\sim$10$^{38}$ erg~s$^{-1}$.  
This $L_{\rm X}$ is consistent, at the high end, with that
of LMC superbubbles \citep{DPC2001}. 
The $Q$(He{\sc ii}) of this diffuse soft X-ray emission is 
at least 4 orders of magnitude lower than that of the hard X-ray point
source.

Contrary to our results, previous studies claim that emission from the
hard X-ray point source in IZw18 may be significant for its He{\sc ii}
ionization \citep[e.g.,][]{l17,heap,S19}. \citet{l17} and
\citet{heap}, however, do not consider the nebular He{\sc ii} budget
in their modeling, which is key to studying the origin of the bulk of
the He{\sc ii} ionizing photons \citep[see, e.g., K15;
K18;][]{kehrig20}.  On the other hand, \citet{S19} assume that the
nebular He{\sc ii} emission in IZw18 is fully ionized by its HMXB,
this way overlooking the energy-dependent cross section of He$^+$
which must be considered for the estimation of the intrinsic X-ray
ionizing power (see above).  Moreover, these works do not
take into consideration the distinct spatial locations of the X-ray
and the He{\sc ii} emissions in IZw18 unveiled here.

\section{Summary and concluding remarks}

Cosmic dawn marks a phase transition of the universe during which
Population III-stars and their hosts put an end to the dark ages. The nature
of these first sources is highly unconstrained. The hard spectra of
Population III and nearly Z=0 stars can produce intense He{\sc ii} emission. However, the
origin of nebular He{\sc ii} emission in metal-poor SF galaxies, near
and far, is still unclear \citep[e.g., K18;][]{saxena}.

IZw18 is a very-metal-poor (Z $\sim$ 3$\%$ Z$_{\odot}$), He{\sc
  ii}-emitting SF system, which is a unique local laboratory for
probing the conditions dominating in distant He{\sc ii} emitters. Here
we combine optical IFS observations with archival HST and X-ray
data of IZw18 to obtain a new view on the effects of X-ray photons
on its nebular He{\sc ii}-ionization.

We confirm the previously reported HMXB which dominates
the IZw18 X-ray emission, although a softer
diffuse emission is also present. 
The observation period covered by the X-ray datasets allows us to study the
X-ray variability of the IZw18 HMXB for the first time.
Its $L_{\rm X,0.5-8~keV}$ is unchanging in hours-days while it can vary by factors of a few in time-scales of years.
There is thus no observational evidence for episodes of highly
  enhanced X-ray emission that, eventually, could power the He{\sc ii} emission in IZw18.

We find that the best-fit Kerr BH and irradiated disk models yield, respectively,
$Q$(He{\sc ii}) $\sim$ 50 and 10 times lower than the observed value.
Using photoionzation grids incorporating both stellar spectra and HMXB
models (S20), we checked that our derived
$log (L_{\rm X,0.5-8~keV}/SFR$) $\sim$ 40.9 corresponds to
He{\sc ii}/H$\beta$ values $\sim$ 9--58 times lower than
the integrated value of He{\sc ii}/H$\beta$ measured in
IZw18 (K16). Finally, we report for the first time a substantial spatial separation ($\approx$200 pc) between the HMXB and He{\sc ii} emission peak.

Although a direct constraint to the $Q$(He{\sc ii}) produced by an HMXB
in extremely metal-poor galaxies, such as IZw18, has not yet been reported,
and one cannot rule out the contribution of X-rays to the He$^{+}$
ionization there, all the points above argue against the HMXB being
the main source responsible for the bulk of the nebular HeII emission
in IZw18. Recent findings obtained by stacking the X-ray emission
  of He{\sc ii}-emitting SF galaxies at z $\sim$ 3, discard X-ray binaries as the dominant producers of He$^{+}$ ionizing photons in
distant SF galaxies \citep[][]{saxena}.  Along with the detailed study of the temporal
variations and high-resolution mapping of the X-ray emission in IZw18
presented here, these studies highlight that the
main source of He{\sc ii}
ionization is still puzzling both in nearby and distant SF systems
(see also K15; K18).


\section*{Acknowledgement}

We thank the referees for helpful comments and suggestions.
This work has been partially funded by 
projects AYA2016-79724-C4-4-P and PID2019-107408GB-C44 from the Spanish PNAYA.
C.K., M.A.G., and J.V.M. acknowledge financial support from the State Agency for Research
of the Spanish MCIU through the ``Center of Excellence Severo Ochoa'' award
to the Instituto de Astrof\'\i sica de Andaluc\'\i a (SEV-2017-0709).  
M.A.G. acknowledges support from MCIU grant PGC2018-102184-B-I00 co-funded with
FEDER funds. G.R.-L. acknowledges support from CONACyT (grant 263373) and PRODEP (M\'exico). This research has made use of data and/or software provided by the High
Energy Astrophysics Science Archive Research Center (HEASARC), which is
a service of the Astrophysics Science Division at NASA/GSFC.

\end{document}